\documentclass[10pt,a4paper,twocolumn]{article}
\usepackage{epsfig}
\begin{document}
\title{Single production of new gauge bosons from the littlest Higgs
model at the $TeV$ energy $e^{-}\gamma$ colliders}

\author{Chongxing Yue and Wei Wang\\
{\small  Department of Physics, Liaoning Normal University, Dalian
116029, China}\thanks{E-mail:cxyue@lnnu.edu.cn}\\}
\date{\today}

\maketitle
\hspace{-0.5cm} {\large\bf Abstract:} In the context of the
littlest Higgs(LH) model, we study single production of the new
gauge bosons $B_{H}$, $Z_{H}$ and $W_{H}^{\pm}$ via $e^{-}\gamma$
collisions and discuss the possibility of detecting these new
particles in the $TeV$ energy $e^{+}e^{-}$ collider($LC$). We find
that these new particles can not be detected via the $e^{-}\nu\nu$
signal in all of the parameter space preferred by the electroweak
precision data. However, the heavy gauge bosons $B_{H}$ and
$Z_{H}$ may be observed via the decay channel
$B_{H}(Z_{H})\rightarrow l^{+}l^{-}$ in wide range of the
parameter space.

\hspace{-0.5cm}PACS number(s):12.60.Cn, 14.70.Pw, 13.66.Hk

\vspace{0.8cm}

It is widely believed that the hadron colliders, such as Tevatron
and future $LHC$, can directly probe possible new physics beyond
the standard model($SM$) up to a few $TeV$, while the $TeV$ energy
linear $e^{+}e^{-}$ collider($LC$) is also required to complement
the probe of the new particles with detailed measurement[1]. A
unique feature of the $TeV$ energy $LC$ is that it can be
transformed to $\gamma \gamma$ or $e \gamma$ collider (photon
colliders) by the laser-scattering method. The effective
luminosity and energy of the the photon colliders are expected to
be comparable to those of the $LC$. In some scenarios, they are
the best instrument for the discovery of signal of new physics.

The $e^{-}\gamma$ collisions can produce particles which are
kinematically not accessible at the $e^{+}e^{-}$ collisions at the
same collider[2]. For example, for the process
$e^{-}\gamma\rightarrow AB$ with light particle $A$ and new
particle $B$, the discovery limits can be much higher than in
other reactions. Furthermore, the initial state photon provides us
with a possibility to directly probe the gauge boson
self-interactions and its cross section is simply dependent on the
coupling parameters, so the $e^{-}\gamma$ collider is particularly
suitable for studying heavy gauge boson production. In this
letter, we will study single production of the heavy gauge bosons
$B_{H}$, $Z_{H}$ and $W_{H}$ predicted by the littlest Higgs($LH$)
model[3] via the $e^{-}\gamma$ collisions and discuss the
possibility of detecting these new particles in the future $LC$
experiments.

Little Higgs models[3,4] were recently proposed as the kind of
models of electroweak symmetry breaking(EWSB), which can be
regarded as the important candidates of new physics beyond the
$SM$. The $LH$ model[3] is one of the simplest and
phenomenologically viable models, which realizes the little Higgs
idea. It consists of a non-linear $\sigma$ model with a global
$SU(5)$ symmetry, which is broken down to $SO(5)$ by a vacuum
condensate $f\sim \Lambda_{s}/4\pi\sim TeV$. The gauged subgroup
$SU(2)_{1}\times U(1)_{1}\times SU(2)_{2}\times U(1)_{2}$ is
broken at the same time to its diagonal subgroup $SU(2)\times
U(1)$, identified as the $SM$ electroweak gauge group. This
breaking scenario gives rise to four massive gauge bosons $B_{H}$,
$Z_{H}$ and $W_{H}^{\pm}$, which might produce characteristic
signatures at the present and future collider experiments[5,6,7].

Global fits to the electroweak precision data produce rather
severe constraints on the parameter space of the $LH$ model[8].
However, if the $SM$ fermions are charged under $U(1)_{1}\times
U(1)_{2}$, the constraints become relaxed. The scale parameter
$f=1\sim 2TeV$ is allowed for the mixing parameters $c$ and $c'$
in the ranges of $0\sim 0.5$ and $0.62\sim 0.73$, respectively[9].

Taking account of the gauge invariance of the Yukawa couplings and
the $U(1)$ anomaly cancellation, the relevant couplings of the
gauge bosons $B_{H}$, $Z_{H}$ and $W_{H}^{\pm}$ to the $SM$
fermions can be written as[5]:
\begin{eqnarray}
g^{B_{H}ll}_{L}&=&\frac{e}{2c_{W}s'c'}(c'^{2}-\frac{2}{5}),\nonumber\\
g^{B_{H}ll}_{R}&=&\frac{e}{c_{W}s'c'}(c'^{2}-\frac{2}{5});\\
g^{Z_{H}ll}_{L}&=&-\frac{e c}{2s_{W}s}, \ \ \ \ \ \
g^{Z_{H}ll}_{R}=0;\\
g^{W_{H}ll'}_{L}&=&-\frac{e c}{\sqrt{2}s_{W}s}, \ \ \ \ \ \
g^{W_{H}ll'}_{R}=0;\\
g^{B_{H}\nu\nu}_{L}&=&\frac{e}{2c_{W}s'c'}(c'^{2}-\frac{2}{5}),\nonumber\\
g^{B_{H}\nu\nu}_{R}&=&0;\\
g^{Z_{H}\nu\nu}_{L}&=&\frac{e c}{2s_{W}s}, \ \ \ \ \ \
g^{Z_{H}\nu\nu}_{R}=0.
\end{eqnarray}

From Eq.(1) we can see that the couplings of the gauge boson
$B_{H}$ to leptons vanish at $c'=\sqrt{2/5}$. Thus, in the
parameter space around $c'=\sqrt{2/5}$, the contributions of
$B_{H}$ to observables are significantly reduced. Ref.[7] has
shown that a large portion of the parameter space($f=1\sim 2TeV,\
c=0\sim 0.5,\ c'=0.62\sim 0.73$) consistent with the electroweak
precision data can accommodate the Tevatron direct searches to
$B_{H}$ decaying into dileptons. The light $B_{H}$ is not excluded
by the direct searches for the neutral gauge boson at the
Tevatron. So, we will take $M_{B_{H}}$ and $M_{Z_{H}}$ in the
ranges of $300GeV\sim 900GeV$ and $1TeV\sim 3TeV$ in our numerical
estimation. The gauge bosons $W_{H}$ and $Z_{H}$ are almost
degenerate in mass, we assume $M_{W_{H}}=M_{Z_{H}}=M$.

In Fig.1 and Fig.2 we plot the single production cross sections of
the new gauge bosons $B_{H}$, $Z_{H}$, and $W^{\pm}_{H}$ via the
processes $e^{+}e^{-}\rightarrow e^{-}\gamma\rightarrow
e^{-}B_{H}$, $e^{+}e^{-}\rightarrow e^{-}\gamma\rightarrow
e^{-}Z_{H}$ and $e^{+}e^{-}\rightarrow e^{-}\gamma\rightarrow
W^{\pm}_{H}\nu_{e}$ as functions of the mixing parameters for
three values of the masses $M_{B_{H}}$ and $M_{Z_{H}}$, the c.m.
energy $\sqrt{s}=1TeV$ and $3TeV$, respectively. From these
figures, we can see that, in most of the parameter space preferred
by the electroweak precision data, the single production cross
sections decrease as the mixing parameters are decreasing. The
single $B_{H}$ production cross section is larger than $30fb$ for
$c'\geq 0.68$ and $M_{B_{H}}\leq 600GeV$. For $c=0.5$ and
$M=1.9TeV$, the single production cross sections of the new gauge
bosons $Z_{H}$ and $W_{H}$ are $49fb$ and $0.78fb$, respectively.

\begin{figure}[htb]
\vspace{-0.5cm}
\begin{center}
\epsfig{file=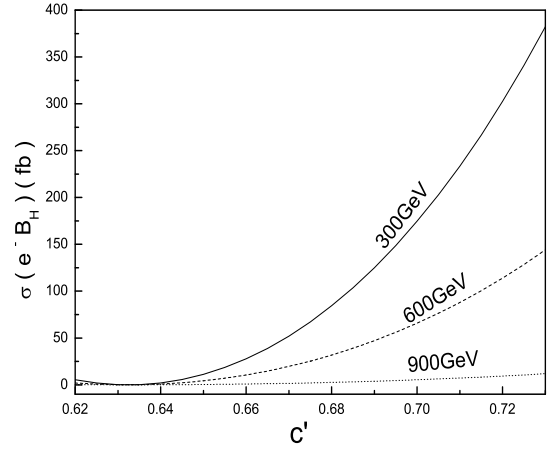,width=230pt,height=200pt} \vspace{-1.25cm}
\hspace{0.5cm} \caption{The single production cross section of
$B_{H}$ as a function of $c'$ for $\sqrt{s}=1TeV$.} \label{ee}
\end{center}
\end{figure}
\vspace*{2.5cm}
\begin{figure}[htb]
\vspace{-3.75cm}
\begin{center}
\epsfig{file=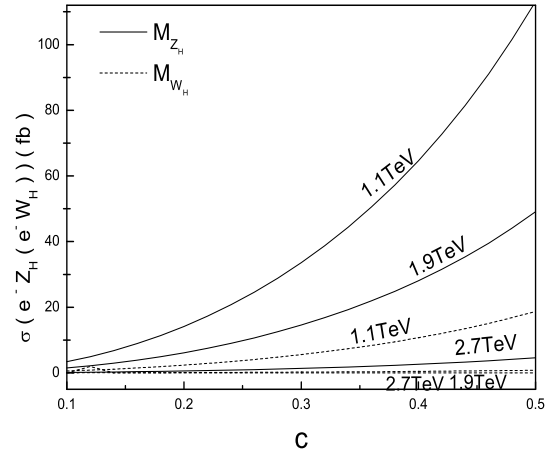,width=230pt,height=200pt} \vspace{-1.25cm}
\hspace{0.5cm} \caption{The single production cross sections of
$Z_{H}(W_{H})$ as functions of $c$ for $\sqrt{s}=3TeV$. Solid and
dashed lines correspond to the gauge bosons $Z_{H}$ and $W_{H}$,
respectively.} \label{ee}
\end{center}
\end{figure}

In general, the heavy gauge bosons are likely to be discovered via
their decays to leptons. If the gauge bosons $B_{H}$ and $Z_{H}$
decay into pair of neutrinos, the signal of the processes
$e^{+}e^{-}\rightarrow e^{-}\gamma\rightarrow
e^{-}B_{H}$($e^{-}Z_{H}$) will be an isolated electron associated
large missing energy. In the narrow width approximation, the
number of observed events can be approximately written as
$N(e\nu\overline{\nu})=\pounds \varepsilon
\sigma(e^{-}B_{H}(e^{-}Z_{H}))\times Br(B_{H}(Z_{H})\rightarrow
\nu\overline{\nu})$, where $\varepsilon$ is the experimental
efficiency for detecting the final state electron and $\pounds$ is
the integrated luminosity of the future $LC$ experiments.
Obviously, in wide range of the parameter space preferred by the
electroweak precision data, the gauge bosons $B_{H}$ and $Z_{H}$
can generate several hundreds and up to thousand
$e\nu\overline{\nu}$ events. For example, if we take
$\pounds=100fb^{-1}$ and $\varepsilon=95\%$, then the
$B_{H}(Z_{H})$ can generate $749(321)$ $e\nu\overline{\nu}$ events
for $M_{B_{H}}(M_{Z_{H}})=600GeV(1600GeV)$ and the mixing
parameter $c'(c)=0.7(0.4)$.

The main backgrounds for the $e\nu\overline{\nu}$ signal come from
the $SM$ resonant processes $e^{-}\gamma\rightarrow
e^{-}Z\rightarrow e^{-}\nu\overline{\nu}$ and
$e^{-}\gamma\rightarrow W^{-}\nu \rightarrow
e^{-}\nu_{e}\overline{\nu}_{e}$. The scattered electrons in the
process $e^{-}\gamma\rightarrow e^{-}Z$ has almost same energy
$E_{e}\approx \sqrt{s}/2$ for $\sqrt{s}\gg m_{Z}$. This process
could be easily distinguished from the signal. Furthermore, the
cross section of the process $e^{-}\gamma\rightarrow e^{-}Z$
decreases as $\sqrt{s}$ increasing, while the cross section of the
process $e^{-}\gamma\rightarrow W^{-}\nu_{e}$ is approaching a
constant at high energies. Thus, the most serious background
process will be $e^{-}\gamma\rightarrow W^{-}\nu_{e} \rightarrow
e^{-}\nu_{e}\overline{\nu}_{e}$. To compare the signal with
background and discuss the possibility of detecting these new
particles, we calculate the ratio of signal over square root of
the background($S/\sqrt{B}$) and find that, for the gauge bosons
$B_{H}$ and $Z_{H}$, the value of $S/\sqrt{B}$ is smaller than $2$
in most of the parameter space allowed by the precision
electroweak constraints. Thus, the gauge bosons $B_{H}$ and
$Z_{H}$ can not be observed via the decay channel
$B_{H}(Z_{H})\rightarrow \nu\overline{\nu}$ in the future
$e^{-}\gamma$ collider.

If the gauge bosons $B_{H}$ and $Z_{H}$ decay to pair of charged
leptons, then the signal of the processes $e^{-}\gamma\rightarrow
e^{-}B_{H}$ and $e^{-}\gamma\rightarrow e^{-}Z_{H}$ is three jets,
two of the three jets are reconstructed to the new gauge boson
mass. Comparing with the signal $e^{-}\nu\overline{\nu}$, the
signal $e^{-}l^{+}l^{-}$ has small background, which mainly comes
from the $SM$ resonant process $e^{-}\gamma\rightarrow e^{-}
Z\rightarrow e^{-}l^{+}l^{-}$(The contributions of the $SM$
process $e^{-}\gamma\rightarrow e^{-}l^{+}l^{-}$ to the background
can be safely neglected at high energy[10]). It is evident that
these new gauge bosons can be easier detected from the
$B_{H}(Z_{H})\rightarrow l^{+}l^{-}$ decay channel than from the
$B_{H}(Z_{H})\rightarrow \nu\overline{\nu}$ decay channel.

In Fig.3 and Fig.4 we plot the $S/\sqrt{B}$ for the signal
$e^{-}l^{+}l^{-}$ as function of the gauge boson masses
$M_{B_{H}}$ and $M_{Z_{H}}$, respectively. As long as the values
of the mixing parameters $c'$ and $c$ are in the ranges of
$0.70\sim 0.73$ and $0.4\sim0.5$, respectively, all values of the
$S/\sqrt{B}$ for the gauge bosons $B_{H}$ with $M_{B_{H}}\leq
650GeV$ and $Z_{H}$ with $M_{Z_{H}}\leq 1700GeV$ are larger than
$5$. Thus, in the sizable parameter space preferred by the
electroweak data, these new gauge bosons should be observed via
detecting the $e^{-}l^{+}l^{-}$ signal in the future $LC$
experiments.

\begin{figure}[htb]
\vspace{-0.5cm}
\begin{center}
\epsfig{file=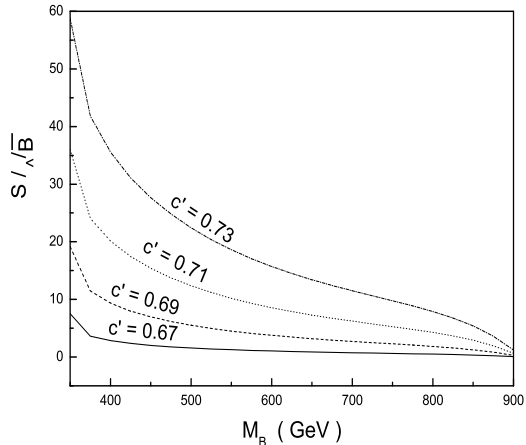,width=230pt,height=200pt} \vspace{-1.5cm}
\hspace{0.5cm} \caption{The $S/\sqrt{B}$ for the signal
$e^{-}l^{+}l^{-}$ as a function of $M_{B_{H}}$ for four values of
the mixing parameter $c'$. } \label{ee}
\end{center}
\end{figure}
\vspace*{1.5cm}
\begin{figure}[htb]
\vspace{-3cm}
\begin{center}
\epsfig{file=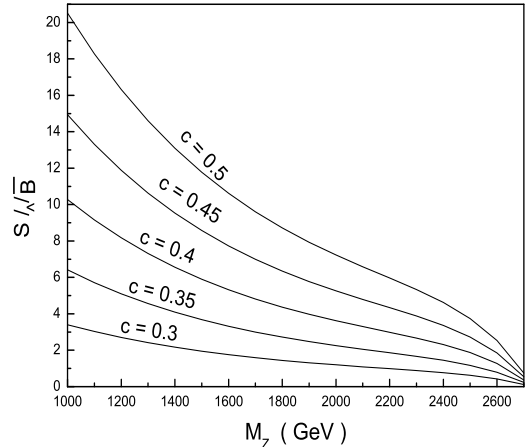,width=230pt,height=200pt} \vspace{-1.5cm}
\hspace{0.5cm} \caption{The $S/\sqrt{B}$ for the signal
$e^{-}l^{+}l^{-}$ as a function of $M_{Z_{H}}$ for five values of
the mixing parameter $c$. } \label{ee}
\end{center}
\end{figure}

The decay $W_{H}^{\pm}\rightarrow l\nu$ can manifest itself via
events that contain an isolated charged lepton and missing energy.
The signal of single production of the gauge bosons $W_{H}^{\pm}$
in the $TeV$ $e^{-}\gamma$ collider should be an isolated charged
lepton associated with large missing energy. However, the
background coming from the process $e^{-}\gamma\rightarrow
W\nu_{e}\rightarrow l\nu \nu_{e}$ is very large and makes that the
value of $S/\sqrt{B}$ is smaller than $2$ in most of the parameter
space preferred by the electroweak precision data. Thus, we have
to say that the new gauge boson $W_{H}$ can not be detected in the
future $e^{-}\gamma$ collider.

The $TeV$ $e^{-}\gamma$ collider is particularly suitable for
studying single production of the heavy gauge bosons. In the
context of the $LH$ model, we have studied the possibility of
detecting the heavy gauge bosons $B_{H}$, $Z_{H}$ and
$W_{H}^{\pm}$ via $e^{-}\gamma$ collision in the future $LC$
experiment with the integrated luminosity $\pounds=100fb^{-1}$. We
find that these new particles can be significantly produced in
wide range of the parameter space preferred by the electroweak
precision data. The gauge boson $B_{H}$ should be observed via
detecting the $e^{-}l^{+}l^{-}$ signal, except for the mixing
parameter $c'\approx \sqrt{2/5}$. It is very difficult to detect
the gauge boson $Z_{H}$ via the $Z_{H}\rightarrow
\nu\overline{\nu}$ decay channel, but it is possible to detect
$Z_{H}$ via the $Z_{H}\rightarrow l^{+}l^{-}$ channel. With
reasonable value of the free parameters, the value of $S/\sqrt{B}$
for $Z_{H}$ can reach 19. Certainly, a more detailed study of
$e^{-}\gamma$ collisions with polarized beams is needed, in order
to enhance the possibility of detecting these heavy gauge bosons
and to distinguish between the $LH$ model and other specific
models beyond the $SM$.

\vspace{1.0cm} \noindent{\bf Acknowledgments}

This work was supported in part by the National Natural Science
Foundation of China under the grant No.90203005 and No.10475037
the Natural Science Foundation of the Liaoning Scientific
Committee(20032101).

\null

\begin{thebibliography}{99}
\bibitem{1}
T. Abe et al. [American Linear Collider Group], {\em
hep-ex}/{\bf0106057}; J. A. Aguilar-Saavedra et al. [{\it
ECFA/DESY } Physics Working Group], {\em hep-ph}/{\bf 0106315}; K.
Abe et al., [{\it ECFA} Linear Collider Working Group
Collaboration], {\em hep-ph}/{\bf 0109166}.
\bibitem{2}
E. Boos et al., {\em  Nucl. In strum. Methods A}{\bf
472}(2001)100; B. Badelek et al. [{\it ECFA/DESY } Photon collider
Working Group], {\em hep-ex}/{\bf0108012}; S. J. Brodsky, {\em
Intern. J. of Mod. Phys. A}{\bf 18}(2003)2871.
\bibitem{3}
N. Arkani-Hamed, A. G. Cohen, E. Katz, A. E. Nelson, {\em JHEP}
{\bf 0207}(2002)034.
\bibitem{4}
N. Arkani-Hamed, A. G. Cohen and H. Georgi, {\em  Phys. Lett.
B}{\bf 513}(2001)232; N. Arkani-Hamed, A. G. Cohen, T. Gregoire
and J. G. Wacker, {\em JHEP} {\bf 0208}(2002)020; N. Arkani-Hamed,
A. G. Cohen, E. Katz, A. E. Nelson, T. Gregoire and J. G. Wacker,
{\em JHEP} {\bf 0208}(2002)021; I. Low, W. Skiba and D. Smith,
{\em Phys. Rev. D}{\bf 66}(2002)072001; M. Schmaltz, {\em Nucl.
Phys. Proc. Suppl.} {\bf 117}(2003)40; D. E. Kaplan and M.
Schmaltz, {\em JHEP} {\bf 0310}(2003)039.
\bibitem{5}
T. Han, H. E. Logan, B. McElrath and L.
T. Wang, {\em Phys. Rev. D}{\bf 67}(2003)095004.
\bibitem{6}
G. Burdman, M. Perelstein and A. Pierce, {\it Phys. Rev. Lett.}
{\bf 90} (2003) 241802. T. Han. H. E. Logen, B. McElrath and L. T.
Wang, {\it Phys. lett. B}{\bf 563}(2003)191; Chongxing Yue,
Shunzhi Wang, Dongqi Yu, {\it Phys. Rev. D}{\bf 68}(2003)115004,
G. Azuelos, et al., {\it hep-ph}/{\bf 0402037}; H. E. Logan. {\it
hep-ph}/{\bf 0405072}; G. A. Gonzalez-Sprinberg, R. Martinez, and
J. Alexis Rodriguez, {\it hep-ph}/{\bf 0406178}; Gi-Chol cho and
Aya Omete, {\it hep-ph}/ {\bf 0408099}; Jaeyong Lee, {\it
hep-ph}/{\bf 0408362}; S. Chang, H.-J. He, {\it Phys. lett. B}{\bf
586}(2004)95.
\bibitem{7}
S. C. Park and J. Song, {\it
Phys. Rev. D}{\bf 69}(2004)115010.
\bibitem{8}
J. L. Hewett, F. J. Petriello and T. G. Rizzo, {\em JHEP} {\bf
0310}(2003)062. C. Csaki, J.Hubisz, G. D. Kribs, P. Meade, and J.
Tering, {\it Phys. Rev. D}{\bf 67}(2003)115002; Chongxing Yue, Wei
Wang, {\it Nucl. Phys. B}{\bf 683}(2004)48; Mu-Chun Chen, {\it
Phys. Rev. D}{\bf 70}(2004)015003.
\bibitem{9}
C. Csaki, J.Hubisz, G. D.
Kribs, P. Meade, and J. Tering, {\it Phys. Rev. D}{\bf
68}(2003)035009. T. Gregoire, D. R. Smith and J. G. Wacker, {\it
Phys. Rev. D}{\bf 69}(2004)115008.
\bibitem{10}
F. M. Renard, {\it Z. Phys. C}{\bf 14}(1982)209; F. Cuypers, G. J.
Van Oldenborgh, R. Ruckl, {\it Nucl. Phys. B}{\bf 383}(1992)45.
\end{thebibliography}
\end{document}